\documentclass{article}
\usepackage{emulateapj} %%% ver. of 18 Sep 96
\usepackage{epsfig}
\usepackage{times}

\newcommand{\one}{{\sc i}}
\newcommand{\two}{{\sc i$\!$i}}
\newcommand{\NR}{\mbox{\scriptsize\sc NR}}
\newcommand{\ER}{\mbox{\scriptsize\sc ER}}
\makeatletter
\@ifundefined{chapter}{\def\thebibliography#1{
\section*{References}
  \list
  {\relax}{\setlength{\labelsep}{0em}
        \setlength{\itemindent}{-\bibhang}
        \setlength{\itemsep}{\parskip}
        \setlength{\parsep}{0pt}
        \setlength{\leftmargin}{\bibhang}}
    \def\newblock{\hskip .11em plus .33em minus .07em}
    \sloppy\clubpenalty4000\widowpenalty4000
    \sfcode`\.=1000\relax}}%
{\def\thebibliography#1{
  \list
  {\relax}{\setlength{\labelsep}{0em}
        \setlength{\itemindent}{-\bibhang}
        \setlength{\itemsep}{\parskip}
        \setlength{\parsep}{0pt}
        \setlength{\leftmargin}{\bibhang}}
    \def\newblock{\hskip .11em plus .33em minus .07em}
    \sloppy\clubpenalty4000\widowpenalty4000
    \sfcode`\.=1000\relax}}

\newlength{\bibhang}
\makeatother

\renewcommand\subtitle{
  \vspace*{-12mm}
  \noindent
 {\scriptsize {\it To appear in} \sc The Astrophysical Journal, 502:\#\#-\#\#, 
     \rm 1998 July 20}\\
  \hbox{\vbox to 9pt{}\tiny Preprint typeset using \LaTeX\ style emulateapj}
}
\lefthead{MOSKALENKO, COLLMAR, \& SCH\"ONFELDER}
\righthead{A COMBINED MODEL OF CYG X-1}

\begin{document}
\hyphenation{brems-strah-lung}

\title{A combined model for the X-ray to $\gamma$-ray emission of Cyg X-1}

\author{Igor V.~Moskalenko\altaffilmark{1}, 
   Werner Collmar, and Volker Sch\"onfelder}

\affil{Max-Planck-Institut f\"ur extraterrestrische Physik,
   Postfach 1603, D-85740 Garching, Germany}

\altaffiltext{1}{and Institute for Nuclear Physics,
   M.V. Lomonosov Moscow State University, 119 899 Moscow, Russia}

%\authoremail{imos, wec, vos@mpe.mpg.de}
\begin{abstract}

We use recent data obtained by three (OSSE, BATSE, and COMPTEL) of four
instruments on board the Compton Gamma Ray Observatory, to construct a
model of Cyg X-1 which describes its emission in a broad energy range
from soft X-rays to MeV $\gamma$-rays self-consistently. 
The $\gamma$-ray emission is interpreted to be the result of
Comptonization, bremsstrahlung, and positron annihilation in a hot
optically thin and spatially extended region surrounding the whole
accretion disk.
For the X-ray emission a standard corona-disk model is applied. We show
that the Cyg X-1 spectrum accumulated by the CGRO instruments during a
$\sim$4 year time period between 1991 and 1995, as well as the HEAO-3
$\gamma_1$ and $\gamma_2$ spectra can be well represented by our
model.  The derived parameters match the observational results obtained
from X-ray measurements.

\end{abstract}

\section{Introduction}
%######################################################################
One of the brightest sources in the low-energy $\gamma$-ray sky, Cyg
X-1, has been extensively studied during the last three decades since
its discovery (\cite{Bowyer65}, for a review see
\cite{Oda77,LiangNolan84}). It is a high-mass binary system
(HDE~226868) with an orbital period of 5.6~days consisting of a blue
supergiant and presumably a black hole (BH) with a mass in excess of
$5M_\odot$ (\cite{Dolan92}). The separation of the two components is
$\approx4\times10^{12}$ cm (\cite{Beall84}). A periodicity of 294~d
found in X-ray and optical light curves is thought to be related to
precession of the accretion disk (\cite{Priedhorsky83,Kemp83}).

The X-ray flux of Cyg X-1 varies on all observed timescales down to a
few milliseconds (e.g., \cite{Cui97}), but the average flux exhibits
roughly a two-modal behaviour.  Most of its time Cyg X-1 spends in a
so-called `low' state where the soft X-ray luminosity (2--10 keV) is
low. The low-state spectrum is hard and can be described by a power-law
with a photon index of $\sim1.7$ in the 10--150 keV energy band. There
are occasional periods of `high' state emission, in which the spectrum
consists of a relatively stable soft blackbody component and a weak and
variable hard power-law component.  Remarkable is the anticorrelation
between the soft and hard X-ray components (\cite{LiangNolan84}), which
is clearly seen during the transition phases between the two states.

Cyg X-1 is believed to be powered by accretion through an accretion
disk. Its X-ray spectrum indicates the existence of a hot X-ray
emitting and a cold reflecting gas. The soft blackbody component is
thought to consist of thermal emission from an optically thick and cool
accretion disk (\cite{ShakuraSunyaev73,Pringle81,Balucinska95}).  The
hard X-ray part ($\ga10$ keV) with a break at $\sim150$ keV has been
attributed to thermal emission of the accreting matter Comptonized by a
hot corona with temperature from tens to hundred keV
(\cite{SunyaevTitarchuk80,LiangNolan84}). A broad hump peaking at
$\sim20$ keV (\cite{Done92}), an iron K$\alpha$ emission line at
$\sim6.2$ keV with an equivalent width $\sim100$ eV
(\cite{Barr85,Kitamoto90}, see also \cite{Ebisawa96} and references
therein), and a strong iron K-edge (e.g., see
\cite{Inoue89,Tanaka91,Ebisawa92},1996) have been interpreted as
signatures of Compton reflection of hard X-rays off cold accreting
material.

In addition, there have also been sporadic reports of a hard spectral
component extending into the MeV region. The most famous one was the
so-called `MeV bump' observed at a $5\sigma$ level during the HEAO-3
mission (\cite{Ling87}). For a discussion of the pre-CGRO data and
$\gamma$-ray emission mechanisms see, e.g., a review by Owens \&
McConnell (1992). The COMPTEL spectrum accumulated over 15 weeks of
real observation time during the 1991--95 time period shows significant
emission out to several MeV (\cite{McConnell97}), which, however,
remained always by more than an order of magnitude below the MeV bump
reported from the HEAO-3 mission.

The annihilation line search provided only tentative (1.9$\sigma$)
evidence for a weak 511 keV line with a flux of
$(4.4\pm2.4)\times10^{-4}$ photons cm$^{-2}$ s$^{-1}$
(\cite{LingWheaton89}). Recent OSSE observations (\cite{Phlips96})
resulted only in upper limits with values of $\le7\times10^{-5}$
cm$^{-2}$ s$^{-1}$ for a narrow 511 keV line and $\le2\times10^{-4}$
cm$^{-2}$ s$^{-1}$ for a broad feature at 511 keV.

Although an unified view for the X-ray spectra of BH candidates and
their spectral states has yet to be constructed, the qualitative
picture seems to be quite clear.  Current popular models include an
optically thick disk component, a hot Comptonizing region (e.g.,
\cite{Haardt93,Gierlinski97}), and/or an advection-dominated accretion
flow (e.g., \cite{Abramowicz95,NarayanYi95} and references therein).
The spectral changes are probably governed by the mass accretion rate
(e.g., \cite{Chen95,Esin97}).

\begin{deluxetable}{lc}%%%%%%%%%%%%%%%%%%%%%%%%%%%%%%%%%%%%%%%%%%%%%%
\tablecolumns{2}
\footnotesize
\setlength{\tabcolsep}{0.25em}
\tablecaption{ Luminosity of Cyg X-1.  \label{table1}}
\tablewidth{7cm}
\tablehead{
\colhead{Energy band} & \colhead{Luminosity, $10^{36}$erg/s}
}
\startdata
$\geq0.02$ MeV  & $26$ \nl
0.02--0.2 MeV   & $20.5$ \nl
0.2--1 MeV      & $4.8$\nl
$\geq1$ MeV     & $0.6$\nl
\enddata
\end{deluxetable}%%%%%%%%%%%%%%%%%%%%%%%%%%%%%%%%%%%%%%%%%%%%%%%%%%%

This picture, however, provides no explanation for the observed
$\gamma$-ray emission (e.g., McConnell et al. 1997). The hard MeV tail
can not be explained by standard Compton models because they predict
fluxes which are too small at MeV energies, and thus another mechanism
is required. The models developed so far connect the $\gamma$-ray
emission with a compact hot core ($\sim400$ keV or more) in the
innermost part of the accretion disk, which emits via bremsstrahlung,
Compton scattering, and annihilation
(\cite{LiangDermer88,SkiboDermer95}), or with $\pi^0$ production due to
collisions of ions with nearly virial temperature (e.g.,
\cite{KolykhalovSunyaev79,JourdainRoques94}).  Li, Kusunose \& Liang
(1996) have shown that stochastic particle acceleration via
wave-particle resonant interactions in plasmas ($\sim100$ keV) around the
BH could provide a suprathermal electron population, and is able to
reproduce the hard state MeV tail.  The possibility of Comptonization
in the relativistic gas inflow near the BH horizon has been discussed
by Titarchuk \& Zannias (1998).

We use the recent data obtained by three of four instruments aboard
CGRO to construct a model of Cyg X-1, which describes
its emission in a wide energy range from soft X-rays to MeV
$\gamma$-rays (\cite{Moskalenko97}).  Instead of a compact
(pair-dominated) $\gamma$-ray emitting region, we consider an optically
thin and spatially extended one surrounding the whole accretion disk.
It produces $\gamma$-rays via Comptonization, bremsstrahlung and
positron annihilation. For the X-ray emission the corona-disk model is
retained.

In section 2 we discuss the combined OSSE--BATSE--COMPTEL spectrum of
Cyg X-1. Our model and the inferred results are described in sections
3--4, and the implications are discussed in section 5. The applied
formalism is given in the Appendix.

\section{Observations}
\begin{figure*}%%%%%%%%%%%%%%%%%%%%%%%%%%%%%%%%%%%%%%%%%%%%%%%%%%%%%%%
%\vspace{15mm}
\psfig{file=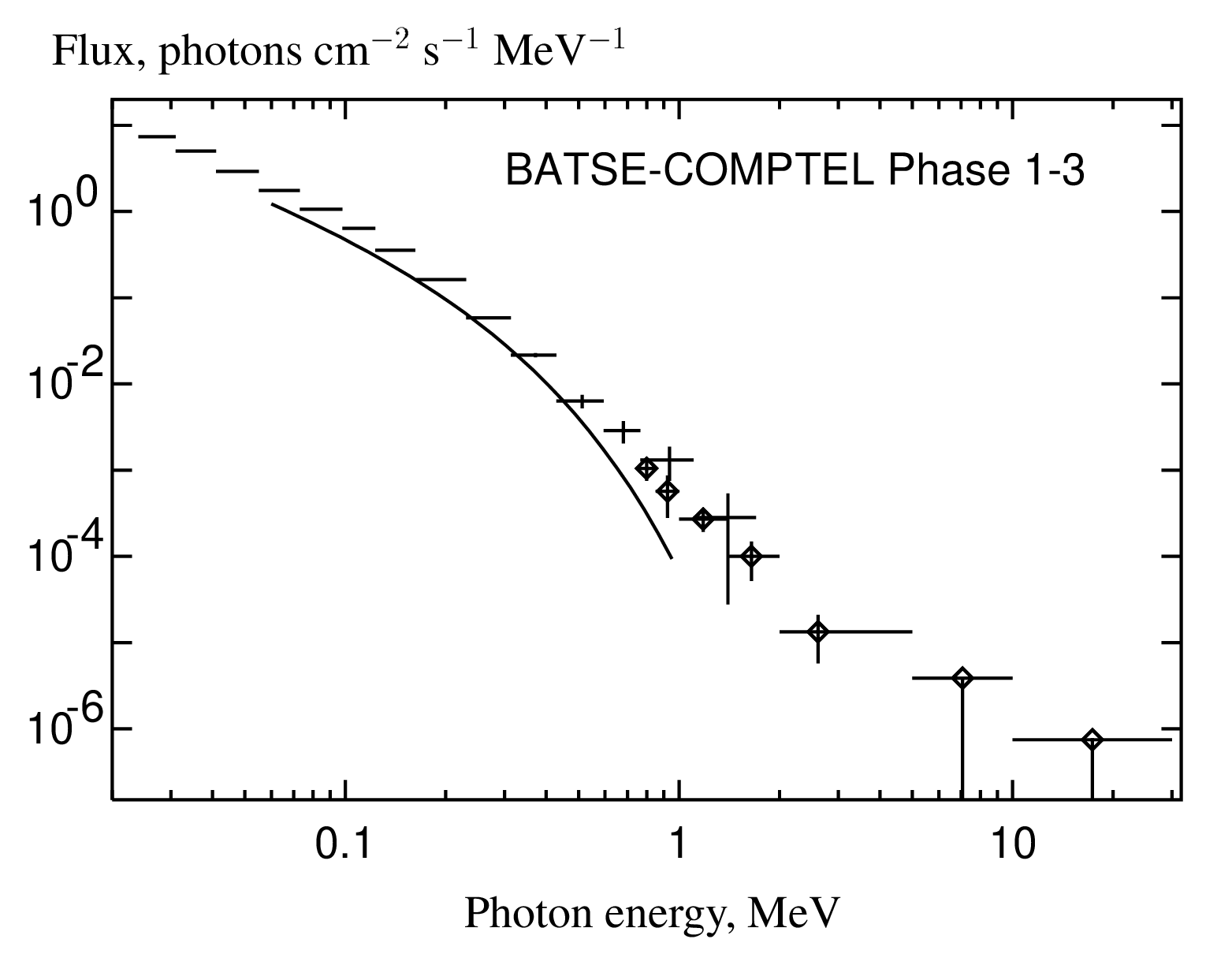,width=9cm}
\hspace{7mm}
\psfig{file=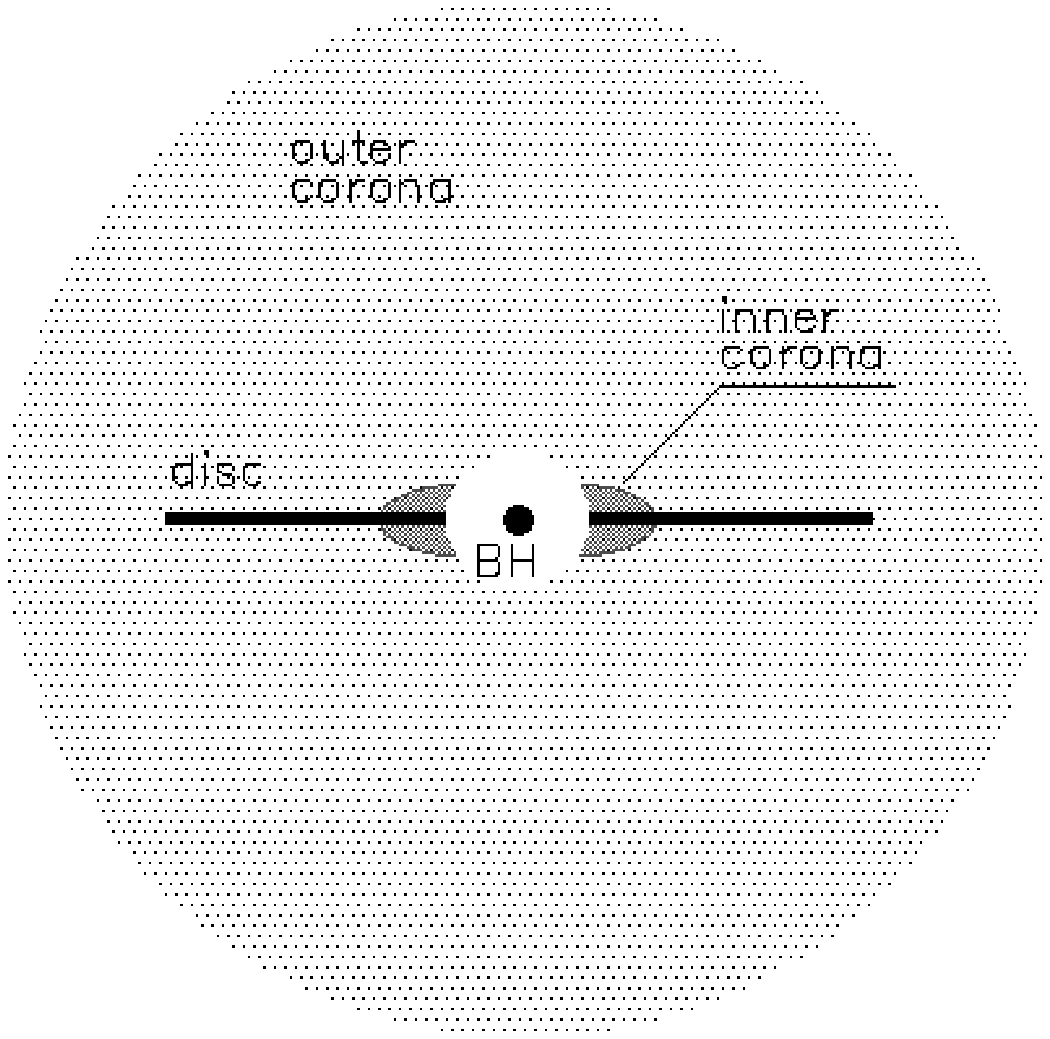,width=6cm}

\parbox{89mm}{%
\caption[fig1.ps]{ Diamonds show the average spectrum of Cyg X-1
based on all COMPTEL CGRO Phase 1--3 observations (\cite{McConnell97})
and crosses represent the almost contemporaneous BATSE data
(\cite{Ling97}). The solid curve represents the best fit to the
time-averaged OSSE spectrum containing 122 days of observation time.
(\cite{Phlips96}).  \label{fig1}}
}\hspace{7mm}
\parbox{89mm}{%
\caption[fig2.ps]{ A schematic view illustrating the model. 
\vspace{19.5mm}
\label{fig2}}
}
%\vspace{15mm}
\end{figure*}

%######################################################################
Since its launch in 1991 Cyg X-1 has been observed by CGRO several
times. The time averaged COMPTEL spectrum based on all observations
between the CGRO Phases 1 and 3 (April 1991 to November 1994) is shown
in Fig.~\ref{fig1} (\cite{McConnell97}) together with the nearly
contemporaneous spectrum derived from BATSE (\cite{Ling97}). The thick
solid curve shows the best fit to the OSSE spectrum (0.06--1 MeV) for
all observations between April 1991 and May 1995 (\cite{Phlips96}). The
best-fit parameters for a power-law model with an exponential cutoff are
a power-law photon index of $\Gamma=1.39$, a cutoff energy $E_c=158$
keV, a normalization intensity of 0.470 photons cm$^{-2}$ s$^{-1}$
MeV$^{-1}$ at 0.1 MeV. The COMPTEL data provide evidence for a hard
power-law tail extending up to at least 3 MeV.

Although the OSSE and BATSE spectra have similar shapes, their intensity
normalizations are different by a factor of $\sim$2 (Fig.~\ref{fig1}).
The discrepancy is largest at the highest energies around 1 MeV. The
COMPTEL measurements lie in between OSSE and BATSE. Although, there is
no way of deducing the exact spectral shape in this region, the total
spectrum is probably smooth, without bumps, which is illustrated by the
three individual spectra.  Possible reasons for this discrepancy have
been discussed by McConnell et al.\ (1997). For our further analysis we
will use the combined BATSE--COMPTEL spectrum.

%%%%%%%%%%%%%%%%%%%%%%%%%%%%%%%%%%%%%%%%%%%%%%%%%%%%%%%%%

Table~\ref{table1} lists the average luminosities of Cyg~X-1 for various
energy bands. The values are derived from the combined BATSE--COMPTEL
spectrum assuming a source distance of 2.5 kpc. The total luminosity
is between 1\% and 10\% of the Eddington luminosity,
\begin{equation}%=====================================================+
\label{2.1}
L_{\rm Edd}\equiv4\pi GM{m_pc\over\sigma_T}\approx1.25\times10^{39}
{\rm\ erg/s\ }\left({M\over10M_\odot}\right).
\end{equation}%=======================================================+

\section{The Model}
%######################################################################
The existence of a compact pair-dominated core around the BH in Cyg X-1
is unlikely in view of the CGRO observations.  The signature of such a
core would be a bump (\cite{LiangDermer88,Liang90}) similar to the one
reported by HEAO-3. However, no evidence for such a bump was detected by
CGRO (\cite{Phlips96,McConnell97}).  In addition, the luminosity of Cyg
X-1 above $\sim0.5$ MeV, though small, would substantially exceed the
Eddington luminosity for pairs, which is $\sim2000$ times lower than
that for a hydrogen plasma.  On the other hand, the hard MeV tail
observed by COMPTEL can not be explained by Comptonization in a corona
of $kT\sim100$ keV and therefore another mechanism is required.

Our study is an attempt to extend the `standard' disk-corona model,
which has been shown to work quite well at X-ray energies (e.g.,
\cite{Gierlinski97,Dove97}), by including the processes of $\gamma$-ray
emission. We investigate the proton-dominated optically thin solution
(\cite{Svensson84}), $\Theta\equiv kT/m_ec^2\la1$, where the
$\gamma$-ray emission is attributed to a spatially extended cloud
surrounding the whole accretion disk (Fig.~\ref{fig2}), the outer
corona, which emits via bremsstrahlung, Comptonization, and positron
annihilation. We concentrate on the hard X-ray to $\gamma$-ray part of
the spectrum, and thus we include into consideration the above-mentioned
processes as well as Comptonization of the soft X-ray disk emission in the
`standard' inner corona. The optical depth of the outer corona has to be
small enough to avoid effective reprocessing of the emission from the
disk and the inner corona. 

The soft X-rays consist of two components,
the local blackbody emission from the disk plus the reflected spectrum.
At energies above $\approx$30 keV the former is negligible and the later
is only of minor importance. Therefore, we neglect both components at
the moment and leave the detailed spectral modelling until the
discrepancy in the intensity normalization of the OSSE and BATSE spectra
has been resolved. However, the effective temperature of the soft excess
is used in calculations of Comptonization in the inner and outer
coronae, and the estimate of the total soft X-ray luminosity is provided
to match the observed value.

%\pagebreak \vfill

Our idea is that the electrons in the outer corona (which is optically
very thin) are relativistic. The heating mechanism is not specified,
but people usually refer to stochastic acceleration (\cite{Li96}), MHD
turbulence in the inner corona (\cite{LiMiller97}), and plasma
instabilities in magnetized advection-dominated accretion flows
(\cite{BisnovatyiLovelace97}). These mechanisms are likely to heat
mainly electrons and so can provide a population of energetic
particles. We further show (see Discussion) that the mean free path of
these electrons in the outer corona is of the same order as its size.
Because there is no mechanism to confine energetic electrons (and pairs
if they exist), except for reasons of charge conservation, they can
move freely inside the outer corona providing the same temperature for
the whole plasma volume. Additionally, the electron cooling in a
thermal plasma at low number density and small optical depth is not
very efficient.

We do not consider the process of $\pi^0$ production in
$pp$-collisions.  Although it could be important at a few Schwarzschild
radii (where the energy of protons is nearly virial), it is unimportant
at tens to hundreds of Schwarzschild radii which is the characteristic
size of the outer corona. The protons in the accreting flow far from
the BH horizon should be cold, since the gravitational forces are quite
weak there and thus the viscous heating in the disk is negligible.
The energy transfer due to the Coulomb coupling with the hot electrons 
is also not efficient.

\subsection{The Fitting Parameters}
%######################################################################
A set of eight fitting parameters was chosen: $kT_i$, $\tau_i$, and
$kT_o$, $\tau_o$, the temperature and optical depth of the inner ({\it
i}) and the outer ({\it o}) coronae, which are assumed to be spheres,
$L^*_{\rm soft}$, the luminosity of the {\it disk} which is {\it
effectively} Comptonized by the inner corona, $L_{\rm soft}$, the {\it
total} effective soft X-ray luminosity of the central source
illuminating the outer corona, $R$, the outer corona radius, and,
$Z=n_+/n_p$, the positron-to-proton ratio in it.

The formulae to calculate the bremsstrahlung, annihililation, and
Comptonization emissivities are given in the Appendix.  The accretion
disk spectrum, which is further reprocessed by the inner
and outer coronae, was
taken to be monoenergetic with an energy $E_0=1.6kT_{bb}$ corresponding
to the maximum of the Planck distribution, where $kT_{bb}=0.13$ keV is
the effective temperature of the soft excess (\cite{Balucinska95}).

The bremsstrahlung and annihilation photon fluxes from the outer corona
are proportional to $R^3n_i n_j$, where $n_{i,j}$ are the number
densities of the plasma particles (see eqs.~[\ref{A.1}],[\ref{A.2}]).  Thus,
if the annihilation contributes significantly, there is a continuum of
solutions given by the set of equations
\begin{eqnarray}%=====================================================+
\label{3.2}
&& R^3n_-n_+\equiv R^3n_p^2 Z(1+Z)=\frac{R\tau_o^2Z(1+Z)}
   {\sigma_T^2(1+2Z)^2}=const, \nonumber\\
&& \tau_o = \sigma_T R n_p (1+2Z) = const,
\end{eqnarray}%=======================================================+
where $kT_o$ is fixed, $n_p$ is the proton number density, 
$\sigma_T$ is the Thomson cross section, and $Z(1+Z)/(1+2Z)^2$ varies 
slowly for $Z\ga0.5$ (therefore, the fitting procedure is not very 
sensitive to this parameter).  If only a negligible positron fraction is 
present, the continuum of solutions is defined by
\begin{equation}%=====================================================+
\label{3.3}
\tau_o = R n_p \sigma_T = const,
\end{equation}%=======================================================+
where $kT_o$ is fixed, $R\leq R_{\rm max}$, and $R_{\rm max}$ is fixed 
from the fitting procedure.

\section{Results}
%######################################################################

\begin{deluxetable}{lcccccc}%%%%%%%%%%%%%%%%%%%%%%%%%%%%%%%%%%%%%%%%%%%%%%
\tablecolumns{7}
\footnotesize
\tablecaption{ The `best-fit' model parameters.  \label{table2}}
%\tablewidth{16.5cm}
\tablehead{
\colhead{} & 
\multicolumn{2}{c}{CGRO Phase 1--3} &
\multicolumn{2}{c}{HEAO-3: $\gamma_2$-state} &
\multicolumn{2}{c}{$\gamma_1$-state} \\
\cline{2-3} \cline{4-5} \cline{6-7} \\
\colhead{Parameters} &
\colhead{\one} & \colhead{\two} &
\colhead{\one} & \colhead{\two} &
\colhead{\one\tablenotemark{a}} & \colhead{\two} }
\startdata

Soft X-ray luminosity, $L_{\rm soft}$ ($10^{36}$ erg s$^{-1}$)
                        & 9.0  &  8.0 & 10.6 & 10.7  & 9.8  & 7.9 \nl
{\it i}-corona temperature, $kT_i$ (keV) 
                        & 76.7 & 79.7 & 95  & 94.9  & \nodata & 93.0 \nl
{\it i}-corona optical depth, $\tau_i$   
                        & 2.39 & 2.23 & 1.41 & 1.42 & \nodata & 1.44 \nl
$L^*_{\rm soft}$, $10^{36}$ erg s$^{-1}$
                        & 0.73 & 0.84 & 1.96 & 1.95 &\nodata & 0.51 \nl
{\it o}-corona temperature, $kT_o$ (keV)
                        & 396  & 436  & 450  & 448  & 346  & 361 \nl
{\it o}-corona optical depth, $\tau_o$
                        & 0.06 & 0.05 & 0.056 & 0.056 & 0.12 & 0.10 \nl
{\it o}-corona radius, $R$ ($10^8$ cm)\tablenotemark{b}
                        & $\la100$ & $\la100$ & $\la100$ & 150 & 391 & 812 \nl
Positron-to-proton ratio, $Z$\tablenotemark{b}
                        & 0    & 0    & 0    & 1.0  & 1.0  & 0.5 \nl
Proton number density, $n_p$ ($10^{10}$ cm$^{-3}$)\tablenotemark{b}
                        & $\ga900$ & $\ga750$ & $\ga840$ & 187  & 154  & 93 \nl
Accretion disk radius, $R_d$ ($10^8$ cm)
                        & \nodata & \nodata & \nodata & 1    & 1    & 1  \nl
511 keV line flux, 
   $I_a$ ($10^{-5}$ photons cm$^{-2}$ s$^{-1}$)\tablenotemark{c}
                        & 0    & 0    & 0    & 0.18 & 0.15 & 0.04\nl
$\chi^2_\nu$            & 4.0  & 3.9  & 1.4  & 1.4  & 0.9  & 0.9 \nl

\enddata

\tablenotetext{a}{The inner corona is small or even absent at all}
\tablenotetext{b}{For $R$, $n_p$, $Z$ dependence see eqs.~(\ref{3.2}),
   (\ref{3.3})}
\tablenotetext{c}{The narrow annihilation line flux from the disk
   (eq.~[\ref{A.5}]) as calculated for the given $R_d$}

\end{deluxetable}%%%%%%%%%%%%%%%%%%%%%%%%%%%%%%%%%%%%%%%%%%%%%%%%%%%%%%

\begin{figure*}%%%%%%%%%%%%%%%%%%%%%%%%%%%%%%%%%%%%%%%%%%%%%%%%%%%%%%%
\psfig{file=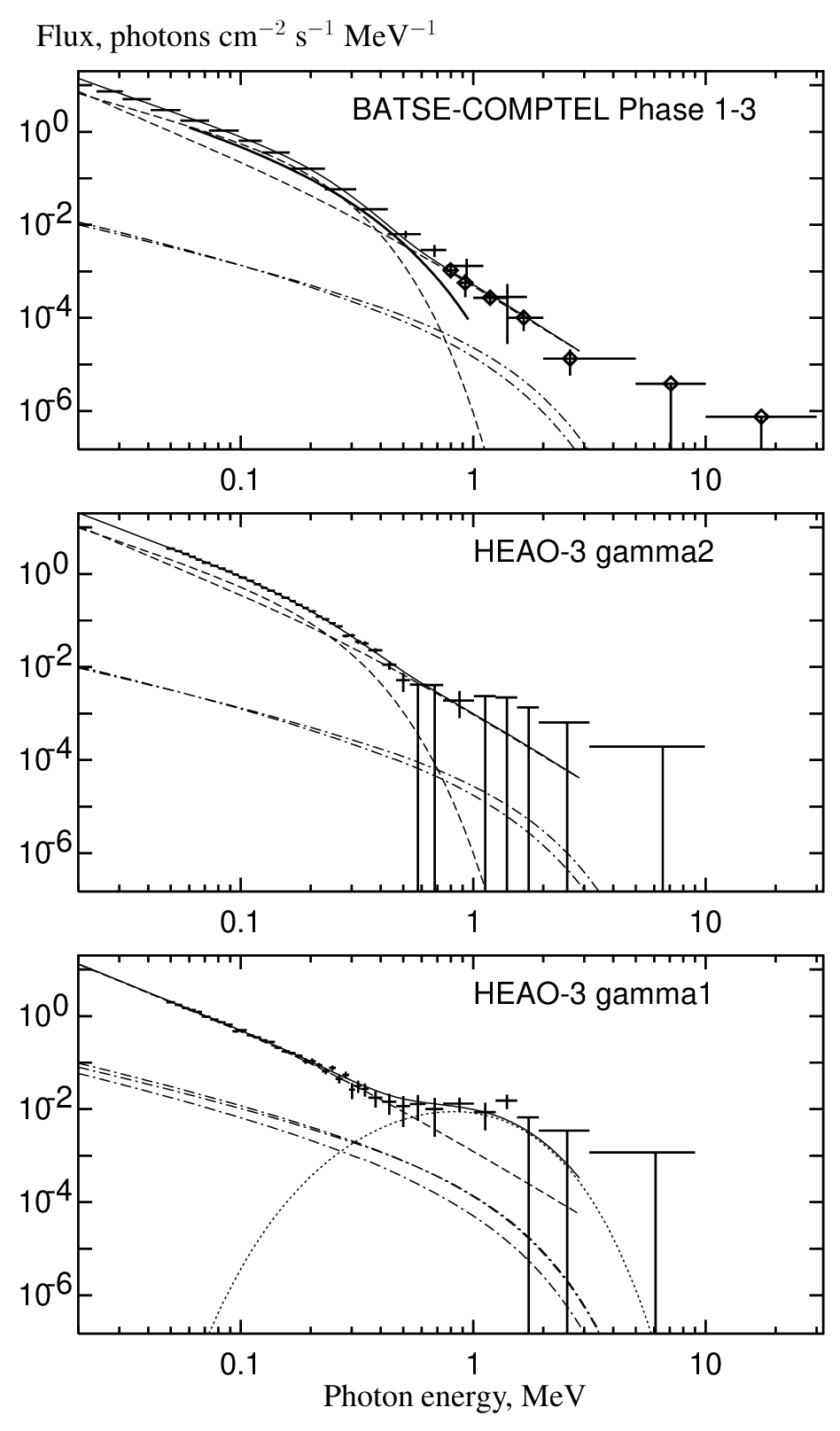,width=9cm}
\hspace{7mm}
\psfig{file=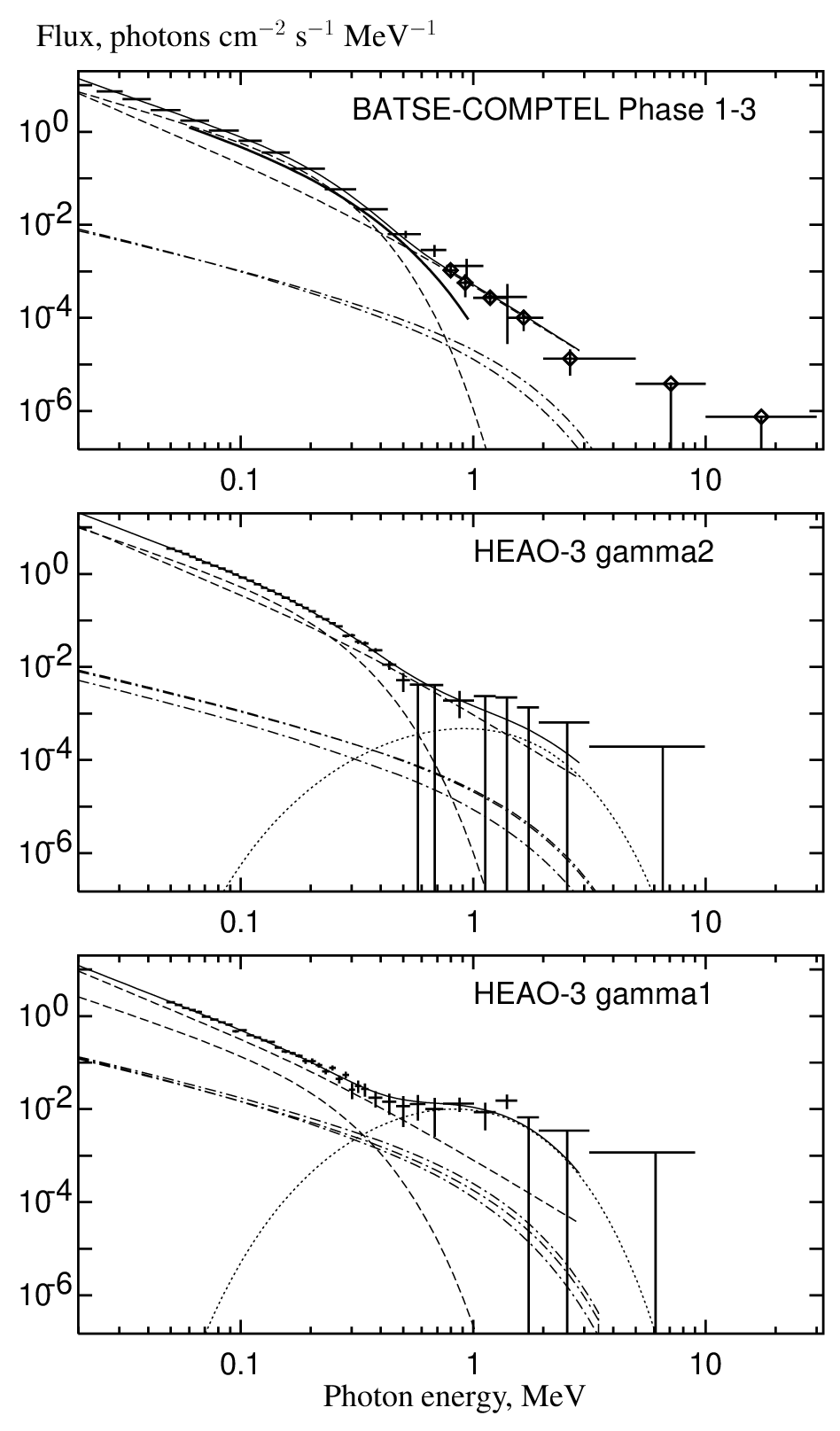,width=9cm}

\parbox{89mm}{%
\caption[fig3.ps]{ {\it Upper panel} shows the calculated Cyg X-1
spectrum together with the data points, which are the same as in
Fig.~\ref{fig1}. {\it Central and lower panels:} the HEAO-3 $\gamma_2$,
and $\gamma_1$ spectra (\cite{Ling87}).  In all panels the thin solid
lines represent our model fit for the parameter sets \one. The
individual spectral components are the annihilation line (dotted line),
$ee$-, $e^+e^-$-, $ep$-bremsstrahlung (dash-dot), and the Comptonized
spectra from the inner and outer coronae (dashed lines). The Comptonized
spectra from the outer corona are shown up to 3 MeV, up to which the
approximation used agrees with Monte Carlo simulations, and where also
significant data points are available. \label{fig3}}}
\hspace{7mm}
\parbox{89mm}{%
\caption[fig4.ps]{ The same as in Fig.~\ref{fig3}, but for parameter
sets \two\ (see Table~\ref{table2}).  \vspace{35mm} \label{fig4}}}
\end{figure*}%%%%%%%%%%%%%%%%%%%%%%%%%%%%%%%%%%%%%%%%%%%%%%%%%%%%%%%%%

The observed spectra of Cyg X-1 are shown in Figs.~\ref{fig3} and
\ref{fig4} together with our model calculations.  The Comptonized
spectrum from the outer corona is only shown up to 3 MeV because of two
reasons: the measurements above consist of upper limits only, and up to
this energy our approximation (see Appendix A.2) has been tested to
agree reasonably with Monte Carlo simulations.  The best-fit parameters
of our model are listed in Table~\ref{table2}.  Two sets of parameters
with the same $\chi^2_\nu$ are shown\footnote{The value of $\chi^2_\nu$
can not be used for the likelihood criterion estimates here mainly
because of the uncertainty in the relative normalization of the OSSE,
BATSE, and COMPTEL data.} for comparison indicating that several
solutions are possible. We consider the first one (\one) however to be
more physical.

The average BATSE--COMPTEL spectrum probably corresponds to the
`normal' (low) state of Cyg X-1. Only two components contribute: the
Comptonized emission from the inner and outer coronae.  Bremsstrahlung
is of minor importance.  By comparing set \one\ and \two\, one can see
that a smaller optical depth of the outer corona corresponds to a
higher temperature. The parameters obtained for the HEAO-3 $\gamma_2$
state are similar, although the spectral upper limits at high energies
($\ga1$ MeV) allow some positron fraction (set \two).

The HEAO-3 $\gamma_1$ `bump' spectrum has not been cofirmed so far, but
if true, it corresponds in our model to an outer corona size which is
several times larger than in the `normal' state, when the inner corona
is small or even absent at all (set \one). A non-negligible positron
fraction (for $R,n_p,Z$ dependence see eq.~[\ref{3.2}]) is too large to
be produced in the optically thin outer corona (\cite{Svensson84}).
Therefore, we suggest a positron production mechanism (i.e., pair
production in $\gamma\gamma$, $\gamma$-particle, or particle-particle
collisions), which might sometimes operate in the inner disk.  The
radiation pressure would necessarily cause a pair wind, which serves as
energy input into the outer corona thereby enlarging its radius.  Note
that matter outflows were found in many accreting binaries. At least
two systems, 1E\,1740.7--2942 and Nova Muscae, provide clear evidence
for pair plasma streams (for a discussion see
\cite{MoskalenkoJourdain97a},b).

A small disk luminosity of $L^*_{\rm soft}\approx10^{36}$ erg/s, which
is Comptonized by the {\it inner} corona, probably implies a geometry
where only the inner part of the disk is effectively covered by the
corona, which means that most of the soft X-ray photons can escape and
reach the observer. The covering factor is estimated to be $\sim$ 0.18
by applying a value of $4.7\times10^{36}$ erg s$^{-1}$ for the total
{\it observed} luminosity of the soft excess (\cite{Balucinska95}, for
a distance of 2.5 kpc). This value agrees well with a covering factor
$\la0.2$ obtained by Dove et al.\ (1997) from self-consistent Monte
Carlo modelling of the corona-disk structure. A slab (plane-parallel)
corona-disk geometry is not capable to reproduce the observed
broad-band X-ray spectrum of Cyg X-1 (\cite{Gierlinski97,Dove97}).

Such a picture is supported by X-ray observations.
The OSSE correlation analysis of source temperature (defined from the
thin thermal bremsstrahlung model) vs.\  45--140 keV intensity
(\cite{Phlips96}) showed that the source temperature and the intensity
vary only within a limited range: $\sim130-170$ keV, and $\sim0.07-0.12$
photons cm$^{-2}$ s$^{-1}$, with few low-temperature -- low-amplitude
exceptions. A similar behaviour of the best-fit bremsstrahlung
temperature vs.\ hard X-ray luminosity (40--200 keV) has been found by
Kuznetsov et al.\ (1997) from the analysis of the entire dataset of the
Granat/SIGMA observations of Cyg X-1 collected between 1990 and 1994.

The soft X-ray ($<10$ keV) luminosity of Cyg X-1 is on the average
$\sim8.5\times10^{36}$ erg/s (e.g., \cite{LiangNolan84,Ebisawa96}).
During the HEAO-3 $\gamma_1$, $\gamma_2$ states it was even lower
(\cite{Ling87}). Taking into account that for the hard X-ray photons
the Comptonization efficiency in the hot plasma drops substantially
(e.g., \cite{HuaTitarchuk95}) and the number of photons decreases as
well, the values of $L_{\rm soft}\approx10^{37}$ erg/s we obtain match
the observational results.

\begin{figure*}%%%%%%%%%%%%%%%%%%%%%%%%%%%%%%%%%%%%%%%%%%%%%%%%%%%%%%%
\psfig{file=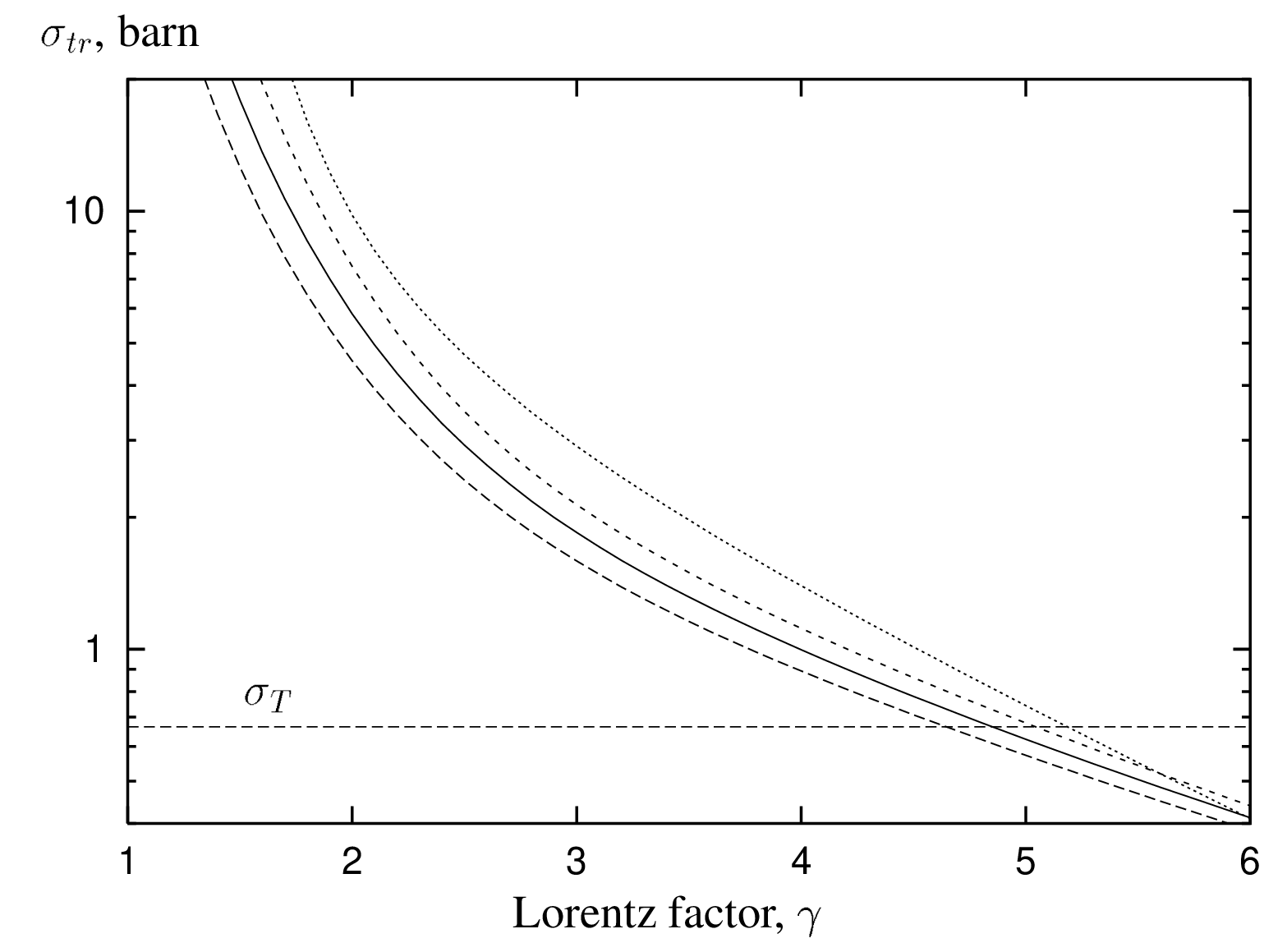,width=9cm} 
\hspace{7mm}
\psfig{file=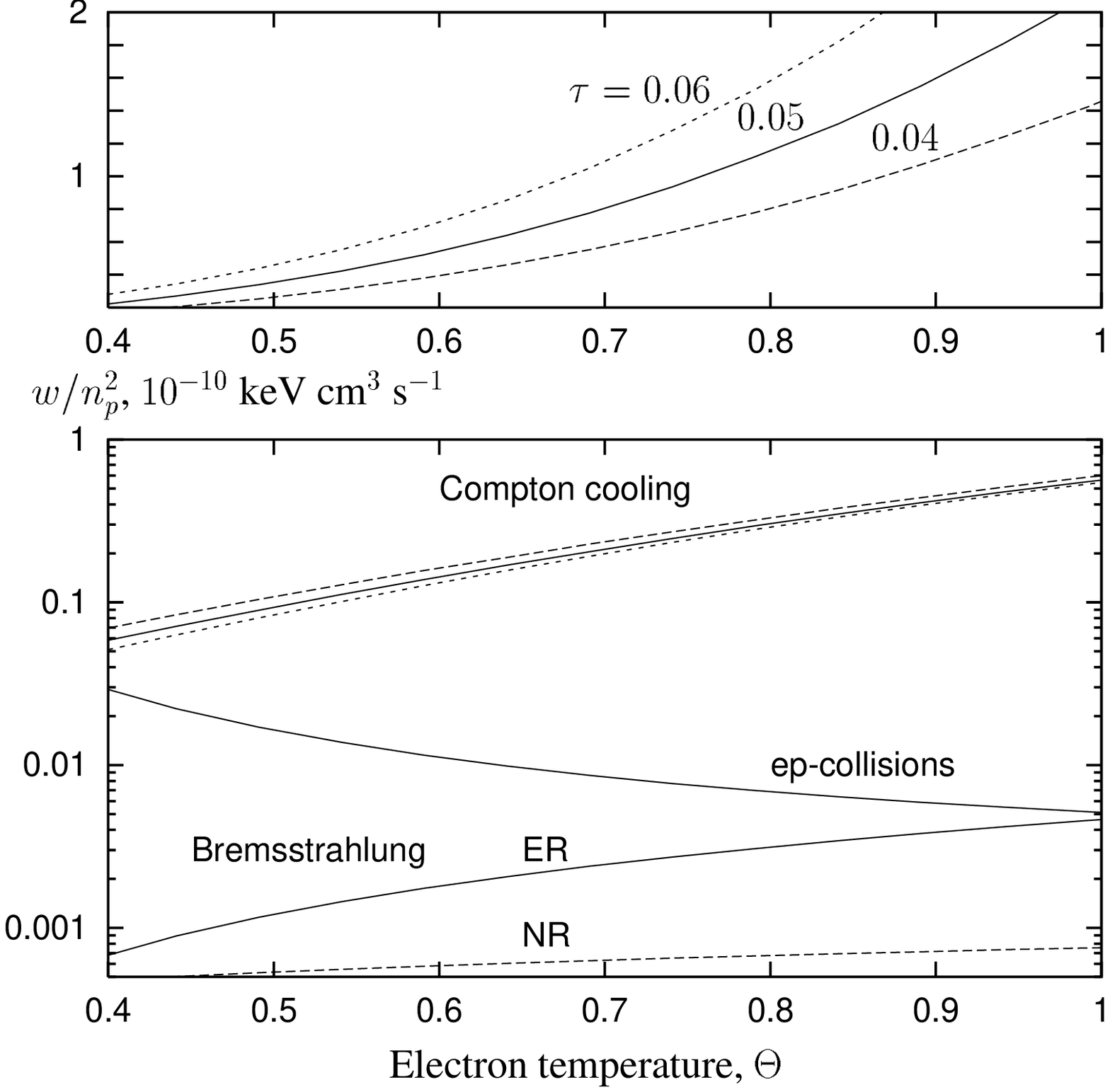,width=9cm}

\vspace{28mm}
\parbox{89mm}{%
\caption[fig5.ps]{ The transport cross section for electrons in a
hydrogen plasma vs.\ the Lorentz factor of a particle. The individual
lines correspond to the plasma temperatures (from top to bottom):
$\Theta=0.2$, 0.6, 0.8, 1.0.  For comparison the Thomson cross section
$\sigma_T$ is also shown. \vspace{15.5mm} \label{fig5}}}
\hspace{7mm}
\parbox{89mm}{%
\caption[fig6.ps]{ {\it Upper panel:} The total cooling rate due to
the Compton scattering as function of the plasma temperature for several
values of the optical depth (eq.~\ref{32}).  {\it Lower panel:} The
cooling rates due to the electron bremsstrahlung $w_{ee}+w_{ep}$ (ER,
NR) and Coulomb coupling with cold protons vs. plasma temperature
($Z=0$, see Appendix). For the Compton scattering shown are the {\it
average} cooling rates calculated for $R=10^{10}$ cm, where the line
styles and the optical depths correspond to these in the upper panel.
\label{fig6}}}
\end{figure*}%%%%%%%%%%%%%%%%%%%%%%%%%%%%%%%%%%%%%%%%%%%%%%%%%%%%%%%%%

No pairs are required to reproduce the spectrum of Cyg X-1 in its
normal state.  If one takes an annihilation line flux of
$I_a\approx4.4\times10^{-4}$ photons cm$^{-2}$ s$^{-1}$
(\cite{LingWheaton89}) in the $\gamma_1$ state, the accretion disk
radius is estimated to be $R_d\sim1.7\times10^9$ cm (eq.~[\ref{A.5}],
set \one). The upper limit allowed by optical measurements is
$R_d\approx6\times10^9$ cm $(M/10M_\odot)$ (\cite{LiangNolan84}), while
the effective radius of the soft X-ray emitting region of the disk is
$\sim4.6\times10^7$ cm (\cite{Balucinska95}).

Our calculations show that the presented model is consistent with the
available observations of the Cyg X-1 system, and is able to reproduce
the observed spectra well. A more detailed study, however, would require
a solution of the discrepancy in the intensity normalization between the
OSSE and BATSE data and further Monte Carlo modelling.

\section{Discussion}
%######################################################################
We have calculated the radiation from Cyg X-1 self-consistently
assuming that the hot optically thin outer corona exists. A mechanism
of its maintenance was not specified in our model (so it is not {\it
totally} self-consistent), but the energy required to maintain such a
corona is quite small and could be provided by a turbulent mechanism,
stochastic particle acceleration,
and/or diffusion of high energy electrons from the inner disk
(e.g., see \cite{Li96} and references therein). 
A relevant example is the solar corona of $\sim10^6$ K (though its
energetic contents is low) compared to 6000 K of the Sun's effective
temperature; but a direct scaling to a BH is not appropriate here.  In
this section we discuss the physical conditions in the outer corona,
i.e. diffusion of electrons and the cooling mechanisms, while do not
touch its origin.

To treate the diffusion of energetic electrons in the outer corona we
consider the {\it transport} cross section for $ee$-scattering. This
allows us (i) to exclude unimportant scattering at small angles
dominating in Coulomb interactions, and (ii) provides us with a correct
estimate of the typical cross section since the $ep$-collisions in a hot
plasma are of minor importance compared to $ee$-collisions.

The transport cross section $\sigma_{tr}$ (see Appendix) for electrons 
in a hydrogen plasma of $\Theta=0.2$, 0.6, 0.8, and 1.0 is shown in 
Fig.~\ref{fig5}.  Although, the value of $\sigma_{tr}$ for particles 
with a Lorentz factor $\gamma\sim2$ is much larger than the Thomson 
cross section, the corresponding mean free path of electrons is close to 
the radius of the outer corona.  This allows electrons to pass freely 
and therefore provide the same temperature for the whole plasma 
volume. Positrons, if produced somewhere, can also homogeneously 
fill the plasma volume.  The annihilation time scale is given by 
(\cite{Svensson82})
\begin{eqnarray}%=====================================================+
\label{13}
&&t_a={1\over \pi r_e^2 c\,n_-}(1+2\Theta^2\ln^{-1}[1.12\Theta+1.3]),
\nonumber\\
&&t_a(\Theta\sim1) \approx400
{\rm\ s}\ \left({10^{12}{\rm\ cm}^{-3}\over n_-}\right).
\end{eqnarray}%=======================================================+
For the parameters listed in Table~\ref{table2} $t_a$ is of the order 
of hundred of seconds.

The relevant cooling rates for electrons in a pure hydrogen plasma
$Z=0$ (see Appendix) are shown in Fig.~\ref{fig6} (lower panel). The
rates are divided by $n_p^2$, the Coulomb logarithm was taken as a
constant of $\ln\Lambda=20$. For comparison we show the {\it average}
Compton energy losses per unit volume (divided by $n_p^2$),
$w_{Cm}n_p^{-2}=W_{Cm}n_p^{-2}\times3/(4\pi R^3) =
W_{Cm}\times3\sigma_T^2/(4\pi R \tau_o^2)$ (see eq.~[\ref{3.3}]),
calculated for $L_{\rm soft}=10^{37}$ erg/s, $\tau_o=0.04$, 0.05, 0.06,
and $R=10^{10}$ cm.  Clearly the {\it average} Comptonization losses
(for $\tau_o$ fixed) depend on the radius of the outer corona and for
$\Theta\ga0.4$ and the parameters adopted substantially dominate
bremsstrahlung losses and losses due to the Coulomb coupling with cold
protons. On the other hand, their total value is not too high, of the
order of $L_{\rm soft}$ (upper panel), which is about 10--20\% of the
total luminosity.

The {\it average} value of the Compton energy losses of an electron in
the outer corona can be estimated as $dE/dt=W_{Cm}/N$, where $N=
4\pi\tau_o R^2/3\sigma_T$ is the total number of electrons in the outer
corona.  Taking the corresponding numerical values
(Table~\ref{table2}), $W_{Cm}\approx L_{\rm soft}\sim 10^{37}$ erg/s,
$\tau_o\sim 0.5$, $R=10^{10}$ cm, one can obtain $dE/dt\approx 200$
keV/s. The appropriate timescale for an electron of $\gamma\sim 2$ is
few seconds which is long compared to $R/c\la$1 sec the particle needs to
cross the outer corona.

\section{Conclusion}
%######################################################################
The data obtained recently by the CGRO instruments allow us to
construct a model of Cyg X-1 which describes its emission from soft
X-rays to MeV $\gamma$-rays self-consistently.  This model is based on
the suggestion that the $\gamma$-ray emitting region is a hot optically
thin and spatially extended proton-dominated cloud, the outer corona.
The emission mechanisms are bremsstrahlung, Comptonization, and
positron annihilation.  For X-rays a standard corona-disk model is
applied.

The CGRO spectrum of Cyg X-1 accumulated over a $\sim$4 years period
between 1991 and 1995, as well as the HEAO-3 $\gamma_1$, and $\gamma_2$
spectra can be well represented by our model.  The derived parameters
match also the basic results of the X-ray observations.  A fine tuning
of the model would require further Monte Carlo simulations and more
accurate spectral measurements. In this respect, the solution of the
discrepancy between the OSSE and BATSE normalizations would be of
particular importance.

\acknowledgements
We thank the referee for useful comments.
Discussions with R.Narayan, L.Titarchuk, and M.Gilfanov are greatly
acknowledged. We are particularly grateful to M.McConnell for providing
us with the combined spectra of Cyg X-1 prior to publication, and
E.Churazov for Monte Carlo simulations of Comptonization in
$\Theta\sim1$, $\tau\approx0.1-0.05$ plasma.

\end{multicols}
\appendix
\section*{\centerline {APPENDIX}}
\section{Radiation from a Thermal Plasma}
%######################################################################
For a thermal plasma consisting of electrons, positrons and protons at
mildly relativistic temperatures ($kT\la m_ec^2$), the main radiation
processes are bremsstrahlung, electron-positron annihilation, and
Compton scattering.

\subsection{Bremsstrahlung}%####
The bremsstrahlung emissivities, the number of photons emitted per unit
time, per unit volume, and per unit energy interval, can be represented
by the form
\begin{equation}%=====================================================+
\label{A.1}
S_{ij}(\varepsilon,kT)\propto n_i n_j{e^{-\varepsilon/\Theta}\over
\varepsilon}G_{ij}(\varepsilon,kT),
\end{equation}%=======================================================+
where $\varepsilon=E/m_ec^2$ is the dimensionless photon energy,
$\Theta=kT/m_ec^2$ is the dimensionless plasma temperature, and
$n_{i,j}$ with $i=\{e^-,e^+\}$, $j=\{e^-, e^+, p\}$ are the
corresponding number densities.  Accurate numerical fits for the Gaunt
factors $G_{ee}(\varepsilon,kT)$ and $G_{e^+e^-}(\varepsilon,kT)$ in an
appropriate energy range have been given by Stepney \& Guilbert (1983)
and Haug (1987), respectively. The $ep$-bremsstrahlung emissivity can
be calculated by the one-fold integration (e.g., see
\cite{StepneyGuilbert83}). The approximations of the Gaunt factors
$G_{ij}(\varepsilon,kT)$ have also been constructed by Skibo et
al.\ (1995).

\subsection{Comptonization}%####
To calculate the effect of Compton scattering in a medium of
$kT\sim100$ keV we follow the model by Sunyaev \& Titarchuk (1980) with
corrections made by Titarchuk (1994) and Hua \& Titarchuk (1995). The total
number of photons emerging from the plasma cloud per unit energy
interval and per unit time is given by
\begin{eqnarray}%=====================================================+
\label{A.3}
F(E,kT)={F_\nu(x,x_0)L_{\rm soft}\over EE_0}, &\ \ x_0\ll1, &
x_0\ll x,
\end{eqnarray}%=======================================================+
where $x\equiv E/kT$, $x_0\equiv E_0/kT$, $E$ is the photon energy,
$E_0$ is the energy of soft photons injected into the plasma, $L_{\rm
soft}$ is the luminosity of the soft photon source, and $F_\nu(x,x_0)$
is the emergent spectrum represented by the Green function
(\cite{HuaTitarchuk95}).

The results of the Hua \& Titarchuk (1995) model are generally in a
good agreement with Monte Carlo simulations except at high
temperatures, $\Theta\sim1$, and small optical depth,
$\tau\sim0.1-0.05$ (\cite{Skibo95}). However, it still provides the
correct spectral index. We found that the disagreement is mainly due to
the steeper tail and the overall normalization, which is overestimated
by the model. A simple power-law with an exponential cutoff,
$\propto(E_0/E)^{\alpha+1} (1-e^{-kT/E})$, where $\alpha$ is determined
by the transcendental equation (\cite{TitarchukLyubarskij95}), gives a
reasonable agreement with simulations up to $\sim3$ MeV.  The
 chosen normalization provides the correct value of the amplification
factor eq.~(\ref{32}).

\subsection{Annihilation}%####
The emissivity of a thermal plasma due to the electron-positron
annihilation is (\cite{Dermer84})
\begin{equation}%=====================================================+
\label{A.2}
S_a(\varepsilon,kT)={n_-n_+c\over kT
K_2^2(1/\Theta)} e^{-{(2x^2+1)\over2x\Theta}}\int_1^\infty d\gamma_r\,
(\gamma_r-1)e^{-{\gamma_r\over2x\Theta}} \sigma_a(\gamma_r),
\end{equation}%=======================================================+
where $K_n$ is the modified Bessel function of the second kind and of 
order $n$, $\gamma_r$ is the relative Lorentz factor of the colliding 
particles (invariant), and $\sigma_a(\gamma_r)$ is the annihilation 
cross section (\cite{JauchRohrlich76}).

The near-Earth intensity of the narrow annihilation line from the disk 
plane can be estimated by the assumption that all positrons which hit 
the disk annihilate in it (two annihilation photons per positron) 
\begin{equation}%=====================================================+
\label{A.5}
I_a\simeq\frac{n_+c}{4}\frac{R_d^2}{D^2} \cos i_d,
\end{equation}%=======================================================+
where $n_+$ is the number density of positrons in the outer corona and 
$\frac{1}4 n_+c$ is the flux density toward the disk surface, $R_d$ is 
the disk radius, $D=2.5$ kpc is the distance, and $i_d$ is the 
inclination angle of the disk plane ($i_d\approx40^\circ$, 
\cite{LiangNolan84}).

\section{Cooling of Electrons}
%######################################################################
The electron cooling in a thermal plasma at low number density and small
optical depth is not very effective. The main channels are:
bremsstrahlung, Comptonization, and Coulomb interactions with ions 
(mainly protons).

\subsection{Bremsstrahlung}%####
For a pure hydrogen plasma the $ep$-bremssrahlung luminosity dominates
the $ee$-bremsstrahlung luminosity in the non-relativistic limit while
at relativistic energies, $\Theta\ga0.5$, the $ee$-bremsstrahlung
dominates. The total energy emitted per {\it unit volume} of plasma
electrons by $ep$- plus $e^+e^-$-bremsstrahlung in the non-relativistic
limit, $\Theta\ll1$, is (\cite{Haug85})
\begin{equation}%=====================================================+
\label{29}
w_{ep}^{\NR}+w_{e^+e^-}^{\NR}\approx{128\over3\sqrt{\pi}}\alpha_f
r_e^2 m_ec^3 n_p^2\sqrt{\Theta}\left({1\over2\sqrt{2}}(1+2Z)+Z(1+Z)\right),
\end{equation}%=======================================================+
where $\alpha_f=1/137$ is the fine structure constant, and $r_e$ is the
classical electron radius.  The total energy emitted by $ee$- plus
$e^+e^-$-bremsstrahlung at the extreme-relativistic energies,
$\Theta\ga1$, is given by (\cite{Alexanian68,Haug85})
\begin{equation}%=====================================================+
\label{30}
w_{ee}^{\ER} =24\alpha_f r_e^2 m_ec^3 n_p^2 (1+2Z)^2
\Theta\{\ln(2\Theta)-0.5772+5/4\},
\end{equation}%=======================================================+
and that for $ep$-bremsstrahlung is (\cite{Stickforth61,Haug75})
\begin{equation}%=====================================================+
\label{31}
w_{ep}^{\ER} =12\alpha_f r_e^2 m_ec^3 n_p^2(1+2Z)
\Theta\{\ln(2\Theta)-0.5772+3/2\}.
\end{equation}%=======================================================+

\subsection{Compton cooling}%####
An expression for the {\it total} energy losses of a plasma volume via
Comptonization has been given by Dermer, Liang, \& Canfield (1991)
\begin{equation}%=====================================================+
\label{32}
W_{Cm} = L_{\rm soft}\frac{P(A-1)}{1-PA}\left[1-
\left(\frac{x_0}{3}\right)^{-1-\ln P/\ln A}\right],
\end{equation}%=======================================================+
where
\begin{eqnarray}%=====================================================+
\label{33}
P &=& 1-e^{-\tau}, \nonumber \\ 
A &=& 1+4\Theta\frac{K_3(1/\Theta)}{K_2(1/\Theta)},
\end{eqnarray}%=======================================================+
$x_0\equiv E_0/kT$ (see eq.~[\ref{A.3}]), and $L_{\rm soft}$ is the 
luminosity of the soft photon source. For $\tau\ll1$ and $x_0\ll1$ 
eq.~(\ref{32}) is almost exact. The luminosity enhancement factor is 
given by $\eta\equiv L/L_{\rm soft}=1+W_{Cm}/L_{\rm soft}$.

\subsection{Coulomb Coupling with Protons}%####
Stepney \& Guilbert (1983) derived a general expression for the 
rate of energy transfer between populations of protons and electrons
with Maxwellian distributions
%\begin{eqnarray}%=====================================================+
\begin{equation}%=====================================================+
\label{34}
w_{Cl} = 4\frac{m_e}{m_p}\pi r_e^2 c\, n_p^2 (1+2Z)\ln\Lambda
\frac{kT_e-kT_p}{K_2(1/\Theta_e)K_2(1/\Theta_p)}
\left[2K_0\left(\frac{\Theta_e+\Theta_p}{\Theta_e\Theta_p}\right)
%\right.
%\nonumber\\
%&&\left.
+\frac{2(\Theta_e+\Theta_p)^2+1}{\Theta_e+\Theta_p}
K_1\left(\frac{\Theta_e+\Theta_p}{\Theta_e\Theta_p}\right)\right],
\end{equation}%=======================================================+
%\end{eqnarray}%=======================================================+
where $\ln\Lambda$ is the Coulomb logarithm, $\Theta_e=kT_e/m_ec^2$, and
$\Theta_p=kT_p/m_pc^2$ are the dimensionless electron and proton
temperatures. The expression is symmetrical with respect to the
electron and proton temperatures.
In the limit of cold protons, $\Theta_p\to0$, eq.~(\ref{34}) reduces to
\begin{equation}%=====================================================+
\label{35}
w_{Cl} = 4\frac{m_e^2}{m_p}\pi r_e^2 c^3\, n_p^2 (1+2Z)\ln\Lambda\,
\frac{2\Theta_e^2 +2\Theta_e +1}{K_2(1/\Theta_e)}.
\end{equation}%=======================================================+

\section{Transport Cross Section}
%######################################################################
Scattering at very small angles dominates in the Coulomb cross section, 
which reflects the long-range nature of the Coulomb interaction. 
However, for the diffusion process in plasma, small scattering angles 
are not very important.  Additionally the $ep$-collisions are of minor 
importance compared to $ee$-collisions.  We therefore restrict ourselves  
by considering the transport cross section only for $ee$-scattering,
which provides us with an estimate on typical values of the relevant 
cross sections.

The transport cross section for a test electron is defined by
\begin{eqnarray}%=====================================================+
\label{B1}
\sigma_{tr}(\gamma_1) &=& \int d^3 p_2 \frac {\sqrt{\gamma_r^2-1}}
{\gamma_1\gamma_2}f(p_2)\int d\Omega\,(1-\cos\theta)
\frac{d\sigma}{d\Omega} \nonumber\\
&=& \int d^3p_2 \frac{ \sqrt{\gamma_r^2-1}}{\gamma_1\gamma_2} f(p_2)
\int d\Omega^*(1-\cos\theta)\frac{d\sigma^*}{d\Omega^*},
\end{eqnarray}%=======================================================+
where $\beta_1$, $\gamma_1$ are the dimensionless speed and the Lorentz
factor of the test particle, $f(p_2)$ is the Maxwell-Boltzmann
distribution, $p_2=\gamma_2\beta_2$ is the momentum of the plasma
particles, $d\sigma/d\Omega$ is the differential cross section of the
Coulomb scattering (\cite{JauchRohrlich76}), and the asterisk marks the
center-of-mass system (CMS) variables.  The scattering angle, 
expressed in CMS variables, is $\cos\theta=(\beta_c+\cos\theta^*)/
(1+\beta_c\cos\theta^*)$, where $\beta_c$,
$\gamma_c=(\gamma_1+\gamma_2)/\sqrt{2(\gamma_r+1)}$ are the speed and
Lorentz factor of the CMS relative to the laboratory system. Changing
to the integration variables $\gamma_2$ and $\gamma_r$ we obtain
\begin{equation}%=====================================================+
\label{B2}
\sigma_{tr}(\gamma_1;\Theta) = {1\over2\beta_1\gamma_1^2\Theta K_2(1/\Theta)}
\int_1^\infty  d\gamma_r (\gamma_r^2-1)^{1/2}
\int_{\gamma^-}^{\gamma^+}
d\gamma_2\, \tilde\sigma_{tr}(\gamma_r,\gamma_c) \,e^{-\gamma_2/\Theta},
\end{equation}%=======================================================+
where $\gamma^\pm=\gamma_1\gamma_r(1\pm\beta_1\beta_r)$, and
\begin{equation}%=====================================================+
\label{B3}
\tilde\sigma_{tr}(\gamma_r, \gamma_c)=\!\!\!
\int_{\cos\theta_M^*}^{\cos\theta_L^*} \!\!\! d(\cos\theta^*)\!
\left[1\! -\! \frac{\beta_c+\cos\theta^*}{1\!
+\!\beta_c\cos\theta^*}\right] \!{d\sigma^*\over d(\cos\theta^*\!)}
\end{equation}%=======================================================+
is the transport cross section for scattering angles $\theta^*$ greater
than the limiting angle $\theta_L^*\to0$, where $\theta_M^*=\pi/2$ for
M{\o}ller scattering and $\theta_M^*=\pi$ for Bhabha scattering.

For M{\o}ller scattering the expression is
\begin{eqnarray}%=====================================================+
\label{B4}
\tilde\sigma_{tr}(\gamma_r, \gamma_c) &=& \frac{4\pi
r_e^2\gamma_r^2}{(\gamma_r-1)^2(\gamma_r+1)}\left[
(4\gamma_c^2-1)\ln\left(\frac{1+\beta_c}{2}\right)
+\frac{2(1-\beta_c)}{1+\beta_c}(\ln\Lambda+\ln\sqrt2)+1 \right]
\nonumber\\ 
&&+\frac1{\gamma_r+1}\left[\left(\frac1{\beta_c^2}-5\right)
\ln(1+\beta_c)-\frac1{\beta_c}+4\ln2+1\right].
\end{eqnarray}%=======================================================+

The mean free path of a test electron in a thermal plasma would be
\begin{equation}%=====================================================+
\label{B5}
\lambda(\gamma_1;\Theta)={\beta_1\over n_-\sigma_{tr}(\gamma_1;\Theta)}.
\end{equation}%=======================================================+

%\clearpage

\end{document}